\documentclass[letterpaper]{article} 
\usepackage{aaai2026}  
\usepackage{times}  
\usepackage{helvet}  
\usepackage{courier}  
\usepackage[hyphens]{url}  
\usepackage{graphicx} 
\usepackage{natbib}  
\usepackage{caption} 
\usepackage{algorithm}
\usepackage{booktabs}
\usepackage{algpseudocode}
\usepackage{afterpage}
\usepackage{hhline}

\usepackage{booktabs}  
\usepackage{array}     
\usepackage{caption}   
\usepackage{inconsolata}
\usepackage{microtype}
\usepackage{multirow}
\usepackage[table]{xcolor} 
\usepackage{booktabs}
\usepackage{makecell}
\usepackage{threeparttable}
\usepackage{graphicx}

\usepackage{enumitem}
\nocopyright 
\usepackage{newfloat}
\usepackage{listings}
\DeclareCaptionStyle{ruled}{labelfont=normalfont,labelsep=colon,strut=off} 
\floatstyle{ruled}
\newfloat{listing}{tb}{lst}{}
\floatname{listing}{Listing}
\title{
SecFSM: Knowledge Graph-Guided Verilog Code Generation for \\ Secure Finite State Machines in Systems-on-Chip
}

\author{
    Ziteng Hu\textsuperscript{\rm 1},
    Yingjie Xia\textsuperscript{\rm 1$^*$},
    Xiyuan Chen\textsuperscript{\rm 1},
    Li Kuang\textsuperscript{\rm 2}
}
\affiliations{
    \textsuperscript{\rm 1}Micro-Electronics Research Institute, Hangzhou Dianzi University; $^*$Corresponding author \\
    \textsuperscript{\rm 2}School of Computer Science and Engineering, Central South University
    \\
    246050116@hdu.edu.cn; xiayingjie@hdu.edu.cn
}

\usepackage{bibentry}
\usepackage{float}
\begin{document}

\maketitle
\begin{abstract}
Finite State Machines (FSMs) play a critical role in implementing control logic for Systems-on-Chip (SoC). Traditionally, FSMs are implemented by hardware engineers through Verilog coding, which is often tedious and time-consuming. Recently, with the remarkable progress of Large Language Models (LLMs) in code generation, LLMs have been increasingly explored for automating Verilog code generation. However, LLM-generated Verilog code often suffers from security vulnerabilities, which is particularly concerning for security-sensitive FSM implementations.
To address this issue, we propose SecFSM, a novel method that leverages a security-oriented knowledge graph to guide LLMs in generating more secure Verilog code. Specifically, we first construct a FSM Security Knowledge Graph (FSKG) as an external aid to LLMs. Subsequently, we analyze users' requirements to identify vulnerabilities and get a list of vulnerabilities in the requirements. Then, we retrieve knowledge from FSKG based on the vulnerabilities list. Finally, we construct security prompts based on the security knowledge for Verilog code generation.
To evaluate SecFSM, we build a dedicated dataset collected from academic datasets, artificial datasets, papers, and industrial cases. Extensive experiments demonstrate that SecFSM outperforms state-of-the-art baselines. In particular, on a benchmark of 25 security test cases evaluated by DeepSeek-R1, SecFSM achieves an outstanding pass rate of 21/25.\footnote{Code will be released upon paper acceptance.}
\end{abstract}

\section{Introduction}

Systems-on-Chip (SoC), a core component of modern Internet of Things (IoT) devices and applications~\cite{saha2024empowering}, has become critical across computing platforms ranging from edge devices such as smartphones~\cite{pohone} and IoT nodes~\cite{iot} to large-scale data centers~\cite{datacenter}. Notably, many SoC modules—such as peripheral controllers, power management units, and bus arbiters—rely on Finite State Machines (FSMs) for state transitions~\cite{socfsmhe,socfsm2019}. 
Traditionally, hardware engineers implement Finite State Machines (FSMs) through manual Verilog coding. This process is often tedious and time-consuming, and crucially, it can lead to security vulnerabilities as SoCs scale up. 
However, the security vulnerabilities in Verilog code for FSM which ultimately undermine their reliability~\cite{attack1,attack2}.

\begin{figure}[!t]
    \centering
    \includegraphics[width=1\columnwidth]{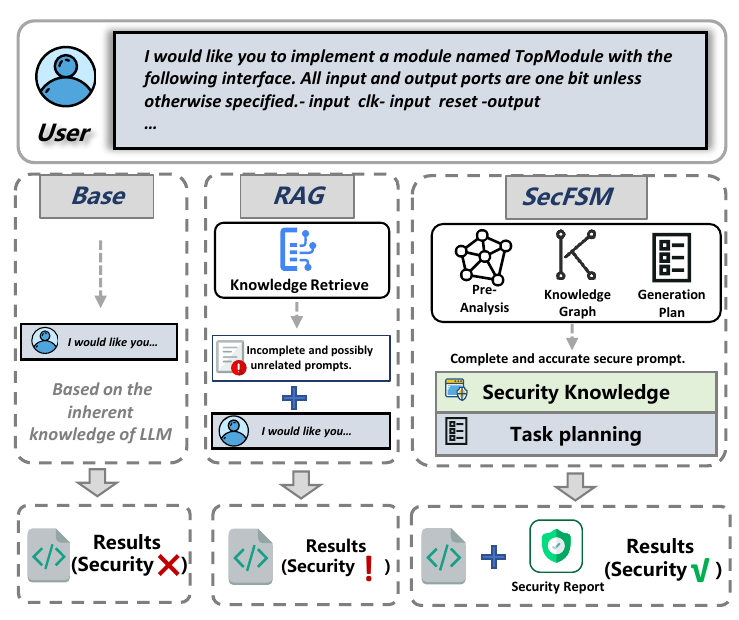} 
    \caption{The motivation of SecFSM. \textit{``Base"} lacks security knowledge. \textit{``RAG"} cannot accurately retrieve knowledge to enhance code security. \textit{SecFSM} uses a collaborative workflow to ensure accurate retrieval and utilization of knowledge.}
    \label{int}
    \vspace{-1em}
\end{figure}

\begin{figure*}[!t]
    \centering
    \includegraphics[width=1\textwidth]{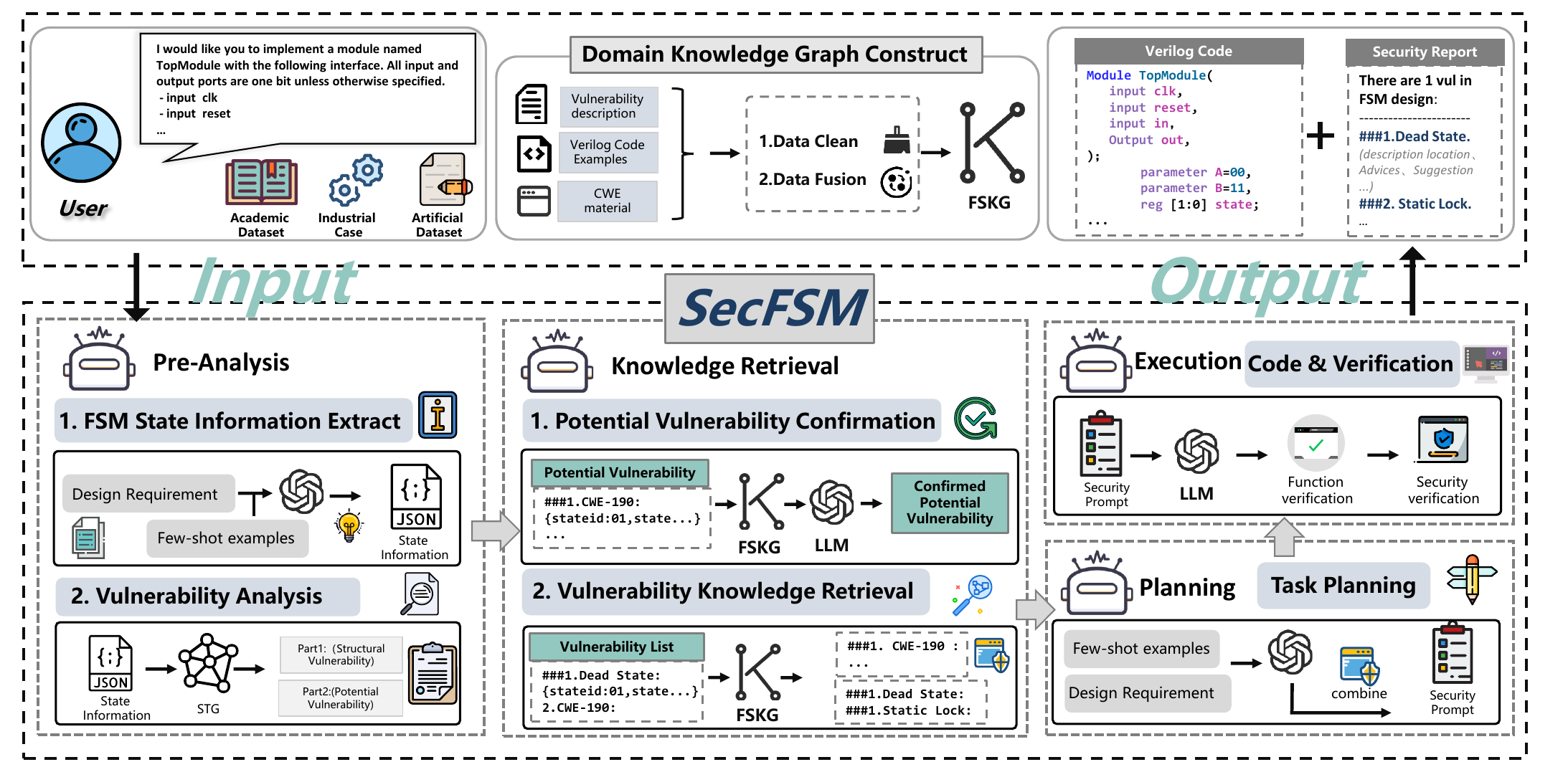}
    \caption{The overall pipeline of SecFSM. We recommend a \textbf{\textit{“Zoom in”}} to view its detailed design: (1)The \textit{pre-analysis} performs pre-analysis base the requirement; (2) The \textit{knowledge retrieval} uses the knowledge of the domain knowledge graph; (3)The \textit{planning} generates a code generation plan and integrates security knowledge. (4)The \textit{execution} generates code and performs verifications.}
    \label{fig:Overview}
\end{figure*}
Recently, Large Language Models (LLMs) have demonstrated remarkable capabilities in diverse fields such as writing and conversational applications, and particularly in Electronic Design Automation (EDA) — including significant advancements in Verilog code generation~\cite{rtlcoder,verilogcoder} and analysis~\cite{chipnemo,verilogeval}.
However, previous works~\cite{verigen} focused on the functionality of generated code, overlooking the security issues of the code.
Although security issues~\cite{saha2024empowering, Don'tCWEAT} related to FSM and Verilog have been studied, LLM often overlooks security issues due to limitations in its knowledge.

Figure \ref{int} shows a motivating example to better illustrate the challenges in the generation of Verilog code for secure FSM.
1) The \textit{``Base"} method indicates that the inherent FSM security knowledge of general LLMs is insufficient to meet the requirements for generating secure Verilog code for FSM.
2) \textit{RAG}, which uses semantic similarity for retrieval, is difficult to accurately retrieve the security knowledge from security knowledge. 
The vulnerability of FSM is in state transition and state information. Therefore, security knowledge cannot be matched by semantic information.
Therefore, the code generated by combining incomplete or irrelevant security knowledge with user requirements is unstable.

To address the aforementioned issues, this paper puts forward a novel method, termed SecFSM, to generate Verilog code for FSM. 
We first build a domain knowledge graph(FSKG). 
Subsequently, we propose a pre-analysis method that combines the state transition structure of FSM to analyze vulnerabilities based on users' requirements. 
Complementing this, we introduce a knowledge graph retrieval method guided by the pre-analysis results. 
Finally, we design a prompt template to guide LLM in Verilog code generation for secure FSM. 

To validate the effectiveness of SecFSM, we conduct extensive experiments on the different datasets, which include academic datasets~\cite{verilogeval}, artificial datasets~\cite{saha2024empowering}, and industrial cases.
Against a benchmark of 25 security test cases assessed by deepseek-r1, SecFSM achieved 21/25, significantly surpassing RAG's 10/25.
In addition, ablation studies have also demonstrated that each module of SecFSM is crucial.

Our primary contributions are as follows:
\begin{itemize}[leftmargin=*, itemsep=0.05mm]
    \item We build a new domain knowledge graph to provide external assistance for LLM, which includes FSM security knowledge.
    \item By developing a novel FSM-based requirement vulnerability pre-analysis, we enable knowledge graph-enhanced multi-stage prompting that guides LLMs to generate secure Verilog code for FSM.d on FSM Design.
    \item  Comprehensive experiments over multi-source datasets validate the effectiveness of our proposed approach, when compared with two baselines. We conduct extensive and holistic ablation studies of each key component.
\end{itemize}

\section{Preliminaries}
\subsection{FSM Knowledge}
\paragraph{FSM.} Finite state machine can be defined as a 6-tuple entity ($S, I, O, s_{0}, \Omega, \gamma$).
According to FSM's characteristics, it can be divided into three parts.
(1) Reset logic. When FSM is reset, it will enter the initial state $s_{0}$.
(2) State transition logic. $\Omega: S \times I \rightarrow S$ is the function that determines the state transition conditions between the states of the FSM.
(3) State output logic.  $\gamma: S \times I \rightarrow O$ is the function that determines the state output logic of Mealy FSM. $\gamma: S \rightarrow O$ is the function of Moore FSM.

\paragraph{Special State.}

A FSM is considered incomplete if the number of states specified in its RTL implementation is less than the maximum allowed by the size of the FSM state register. The unspecified states in such an FSM are designated as don’t-care states.
A Protected State refers to a specific state within a FSM identified as critical from a security perspective. This includes states where security-critical signals are asserted or where unauthorized access or bypass could compromise system security.
Due to the existence of these special states in FSM, proper handling of these states becomes crucial.

\subsection{Problem Formulation}
A typical secure coding task takes functional descriptions and requirements as input and uses an LLM to generate code. However, due to LLMs' lack of domain-specific security knowledge for Verilog, the generated code often overlooks security vulnerabilities. Building on prior work~\cite{verilogcoder,graph}, we propose a graph-based vulnerability pre-analysis method and leverage a domain knowledge graph for remediation guidance. Using pre-analysis results, our approach retrieves relevant security knowledge from the graph to steer the LLM toward generating Verilog code for secure FSM.

Formally, given a natural language query $Q$ about a FSM design requirement and a domain knowledge graph $G_{K}$, the objective of SecFSM is to generate a Verilog code $V_{Se}$ for a secure FSM and a security report $R_{Se}$.

\section{Methodology}
\subsection{An Overview}
Figure \ref{fig:Overview} presents an overview of SecFSM. The method comprises four components: pre-analysis, knowledge retrieval, planning, and execution, which collaboratively function to transform natural language queries into secure Verilog code and a security report. Commencing with a user query $Q$ and knowledge graph $G_{k}$, the pre-analysis method employs graph algorithms to analyze $G_{ST}$ based on $Q$, generating a vulnerability list that encompasses both potential vulnerabilities $V_{P}$ and structural vulnerabilities $V_S$. 
Subsequently, the knowledge retrieval component operates in two distinct phases: (1) confirming potential vulnerabilities $V_P^C$ based on $G_{k}$, and (2) retrieving relevant security knowledge $K_{S}$ and $K_{C}$ from $G_{k}$. This process yields a security report $R$. 
Finally, the execution component generates Verilog code $V_{Se}$ for secure FSM and conducts both functionality and security verification.

\begin{table}[!t]
\small
\centering
\caption{Knowledge Graph Node Relationships}
\captionsetup{skip=2pt}
\label{tab:kg_relationships}
\begin{tabular}{l|l}
\toprule
\textbf{Node Type} & \textbf{Relationships} \\
\midrule
Vulnerability & -- \\
\hline
stage & Vulnerability $\rightarrow$ [:stage] \\
type & Vulnerability $\rightarrow$ [:type] \\
Check & Vulnerability $\rightarrow$ [:Check] \\
Consequence & Vulnerability $\rightarrow$ [:Consequence] \\
GoodExample & Vulnerability $\rightarrow$ [:GoodExample] \\
BadExample & Vulnerability $\rightarrow$ [:BadExample] \\
suggestions & Vulnerability $\rightarrow$ [:suggestions] \\
\hline
manner & alleviation\_suggestions $\rightarrow$ [:manner] \\
confirm & Vulnerability $\rightarrow$ [:confirm] \\
confirm\_positive & confirm $\rightarrow$ [:confirm\_positive] \\
confirm\_negative & confirm $\rightarrow$ [:confirm\_negative] \\
positive\_example & confirm\_positive $\rightarrow$ [:positive\_example] \\
negative\_example & confirm\_negative $\rightarrow$ [:negative\_example] \\
\bottomrule
\end{tabular}
\end{table}
\subsection{Domain Knowledge Graph}
To ensure knowledge consistency, standardization, reliability, and stable knowledge inference, a top-down construction method is adopted to build the FSM Security Knowledge Graph FSKG $G_{K}$.

\paragraph{Ontology Design.}
The FSKG $G_{K}$ is constructed using the Top-Down method, wherein the knowledge graph ontology is defined and presented in Table \ref{tab:kg_relationships}.

\paragraph{FSKG Construction.}

While substantial public knowledge on FSM security is available online, the formats and quality of these documents are highly inconsistent. 
Therefore, constructing a tailored FSKG is essential before deploying our security generation framework. 
The security knowledge comes from relevant papers~\cite{saha2024empowering, Arc-fsm-g}, CWE websites, and related datasets websites. 
Due to the scale of the graph, inconsistencies between knowledge are eliminated through manual screening methods. Then perform entity alignment. Achieving the goal of cleaning and fusing data.

\begin{algorithm}[!t]
\caption{FSM Vulnerabilities pre-analysis}
\label{alg:vuln_Preanalysis}
\begin{algorithmic}[1]
\Require $G_{ST}$: Security State Transition Graph
\Require $\mathcal{V}_{\mathit{struct}}$: Set of structural vulnerability detectors
\Require $\mathcal{V}_{\mathit{pot}}$: Set of potential vulnerability detector
\Ensure ${V}_{S}$,${V}_{{P}}$: Vulnerability list

\State ${V}_{S} \gets \emptyset$
\State ${V}_{{P}}\gets \emptyset$

\Comment{Phase 1: Structural Vulnerability}  
\For{\textbf{each} state $s \in \mathit{G_{ST}}.\mathtt{states}$}
    \For{\textbf{each} detector $\delta \in \mathcal{V}_{\mathit{struct}}$}
        \If{$\delta(s, G_{ST})$}
            \State ${V}_{S} \gets {V}_{S} \cup \{ \mathtt{type}: \delta.\mathit{id}, \mathtt{location}: s \}$
        \EndIf
    \EndFor
\EndFor

\Comment{Phase 2: Potential Vulnerability}  
\For{\textbf{each} state $s \in \mathit{G_{ST}}.\mathtt{states}$}
    \For{\textbf{each} detector $\pi \in \mathcal{V}_{\mathit{pot}}$}
        \If{$\pi(s, G_{ST})$}
            \State ${V}_{P} \gets {V}_{P} \cup \{ \mathtt{type}: \pi.\mathit{id}, \mathtt{location}: s \}$
        \EndIf
    \EndFor
\EndFor

\State \Return ${V}_{S}, {V}_{{P}}$
\end{algorithmic}
\end{algorithm}

\subsection{Pre-analysis}
The pre-analysis comprises two stages:
FSM State Information Extraction ($Q \rightarrow I_{ST}$): State information ($I_{ST}$) is extracted from the requirements ($Q$).
Vulnerability Analysis ($G_{ST} \rightarrow V_S, V_P$): Using the extracted state transition information ($I_{ST}$), an FSM security state transition graph ($G_{ST}$) is constructed. Subsequently, Algorithm \ref{alg:vuln_Preanalysis} is applied to $G_{ST}$ to derive the results of the pre-analysis, namely the structural vulnerabilities ($V_S$) and potential vulnerabilities ($V_P$).

\paragraph{Extract FSM State Information.}
The efficacy of LLMs in extracting FSM security information critically depends on input prompt structure. To ensure accuracy, we designed and implemented a specific few-shot prompt strategy. 
For vulnerability pre-analysis, these requirements are formatted to extract three core elements:
(1) Reset State Information.
(2) State Information (outputs and protection information).
(3) State Transition Information.

\paragraph{Analyze the Vulnerabilities.}

The Algorithm \ref{alg:vuln_Preanalysis} introduces the method for Vulnerability pre-analysis.  It comprises two parts: structural vulnerability identification ($G_{ST}\rightarrow V_{S}$) and potential vulnerability identification ($G_{ST}\rightarrow V_{P}$).
(1) Structural vulnerability $V_{S}$: This phase involves detecting vulnerabilities inherent in the state machine's structure. 
For example, the identification of the ``Dead State". In Algorithm \ref{alg:vuln_Preanalysis}, it is detected by detecting whether there is a ``TRANSITION" relationship entering a state.
(2) Potential Vulnerability  ($G_{ST}\rightarrow V_{P}$): This phase focuses on flagging vulnerabilities that may arise due to specific operations or data handling within states or transitions, based on structural characteristics. 
For example, the analysis of the ``CWE-190". It is related to ``Output" information in $G_{ST}$. Therefore, if the output of a state is conditional (such as the sum of two input parameters), it is considered to have potential vulnerability.vulnerabilities.

\subsection{Knowledge Retrieval}
\paragraph{Potential Vulnerability Confirmation.}
This stage employs vulnerability confirmation knowledge within FSKG ($G_K$) to validate the ``potential vulnerabilities" ($V_{P}$) identified during the pre-analysis phase, resulting in a set of confirmed vulnerabilities ($V_{P}^C$). 
This validation process involves retrieving relevant vulnerability knowledge from $G_K$ and verifying the presence of each $V_P$ instance based on the pre-analysis findings. 
The final output is the confirmed list $V_{P}^C$. For instance, the process verifies whether the buffer overflow vulnerability associated with ``CWE-190" is indeed present.
 
\paragraph{Vulnerability Knowledge Retrieval.}
The security knowledge derived from FSKG ($G_K$) is categorized into two distinct components:
(1) Output security report through vulnerability list $(V_{P}^C,V_{S}\rightarrow K_{S}\rightarrow R)$: Vulnerability arising from design flaws (e.g., ``Dead State"). Notably, automated vulnerability repair utilizing LLMs for such vulnerability may deviate from intended functionality. Consequently, $K_{S}$ is consolidated into a security report ($R$) to provide security recommendations for user reference. 
(2) Output security knowledge for code generation ($V_{P}^C,V_{S}\rightarrow K_{C}$): Vulnerability knowledge preventable through rigorous coding (e.g., ``CWE-1245'' and ``Duplicate Encoding''). $K_{C}$ constitutes an integral element of security prompts.

\subsection{Planning \& Execution}
\subsubsection{Planning.}
The prompt template generation incorporates a foundational component responsible for generating structural elements. The prompt template includes five sections: (1) Input-Output Interface, (2) State Encoding and Associated Declarations, (3) State Transition Logic, (4) State Update Logic and (5) State Output Logic. 
We use the method of few-shot learning to assist LLM in prompt templates $P$.
Subsequently, we concatenate the security knowledge $K_{C}$ and prompt templates $P$ required to generate secure Verilog code to obtain the complete security prompt ($K_{C} + P \rightarrow P_{Se}$).
\subsubsection{Execution.}Finally, LLM generates Verilog code $V_{Se}$ according to $P_{Se}$. Then, carry out the function verification and security verification.

\begin{figure}[h]
    \centering
    \includegraphics[width=0.42\textwidth]{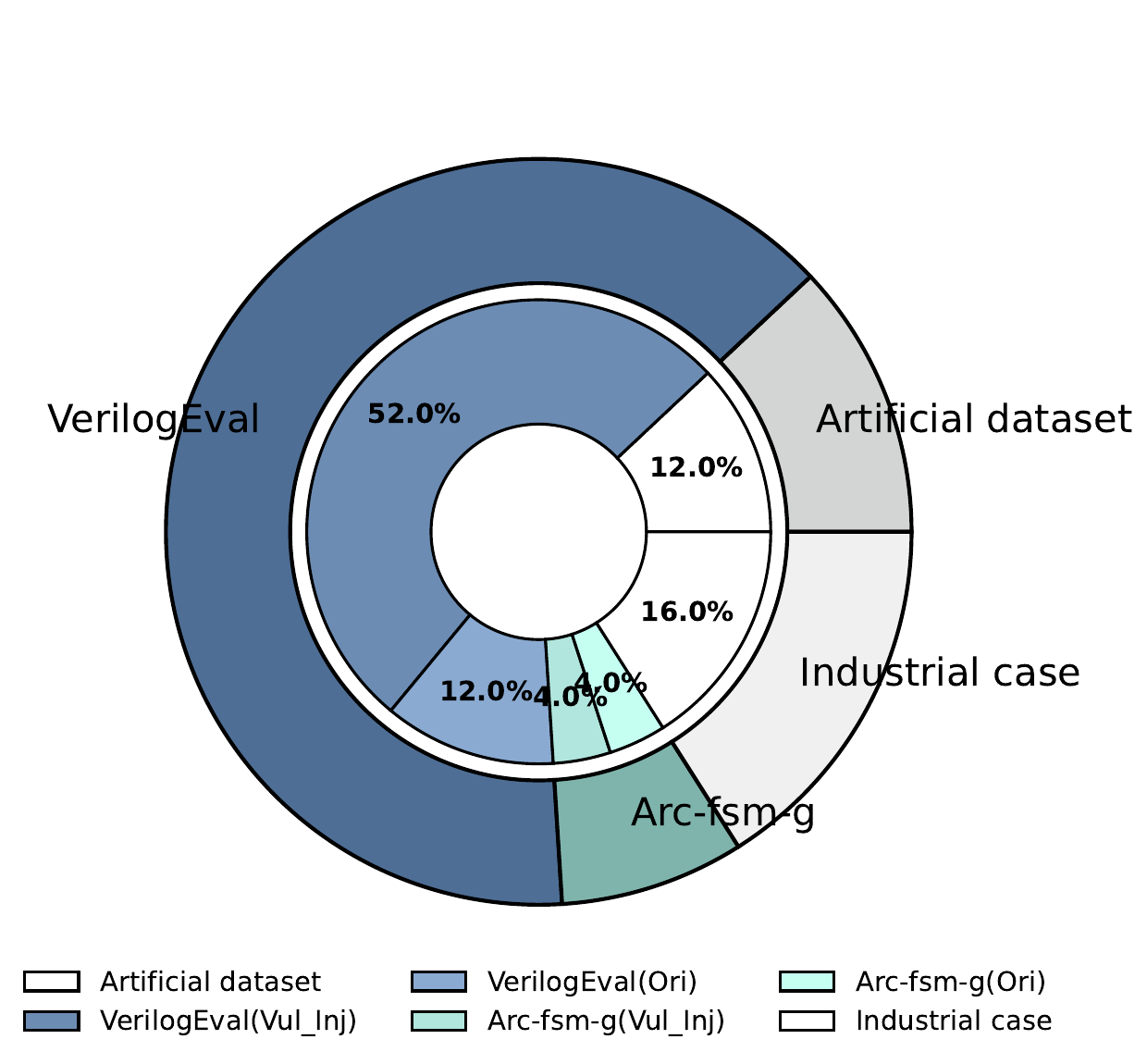}
    \caption{Composition of test data. ``(Ori)" represents a dataset without injected vulnerabilities. ``(Vul\_Inj)" represents the dataset with injected vulnerabilities.}  
    \label{fig:dataset}
\end{figure}
\begin{table*}[!t]
\centering

\begin{threeparttable}
\small
\captionsetup{skip=2pt}
\caption{\label{Security Evaluation results} Overall Performance.}
\setlength{\tabcolsep}{2.3pt} 
\renewcommand{\arraystretch}{1}
\begin{tabular}
{ll||cc|cc|cc||cc|cc|cc||cc|cc|cc}
\hline
  & \textbf{Case}
  & \multicolumn{2}{c|}{\textbf{\makecell{GPT-4o}}} 
  & \multicolumn{2}{c|}{\textbf{\makecell{GPT-4o\\+RAG}}} 
  & \multicolumn{2}{c||}{\textbf{\makecell{SecFSM\\(GPT-4o)}}} 
  & \multicolumn{2}{c|}{\textbf{\makecell{Claude3.5}}} 
  & \multicolumn{2}{c|}{\textbf{\makecell{Claude3.5\\+RAG}}} 
  & \multicolumn{2}{c||}{\textbf{\makecell{SecFSM\\(Claude3.5)}}} 
  & \multicolumn{2}{c|}{\textbf{\makecell{Deepseek-R1}}} 
  & \multicolumn{2}{c|}{\textbf{\makecell{Deepseek-R1\\+RAG}}} 
  & \multicolumn{2}{c}{\textbf{\makecell{SecFSM\\(Deepseek-R1)}}} \\
\cline{3-20}
~ &~ & \textbf{Fun} & \textbf{Sec} 
  & \textbf{Fun} & \textbf{Sec} 
  & \textbf{Fun} & \textbf{Sec} 
  & \textbf{Fun} & \textbf{Sec} 
  & \textbf{Fun} & \textbf{Sec} 
  & \textbf{Fun} & \textbf{Sec} 
  & \textbf{Fun} & \textbf{Sec} 
  & \textbf{Fun} & \textbf{Sec} 
  & \textbf{Fun} & \textbf{Sec} \\
\hline
\multirow{16}{*}{\uppercase\expandafter{\romannumeral1}} 
& q2fsm $^*$ &  0 & -
        &  2 & $\times$
        &  5 \cellcolor{gray!15} & $\times$\cellcolor{gray!15} 
        &  2 & $\times$
        &  3 & $\times$ 
        &  5 \cellcolor{gray!15} & $\times$\cellcolor{gray!15}
        &  5  & $\times$
        &  5  & $\times$
        &  5 \cellcolor{gray!15} & $\sqrt{}$\cellcolor{gray!15}  \\
~ &fsm3s $^*$  &  0 & -
        &  5 & $\times$
        &  5 \cellcolor{gray!15} & $\times$\cellcolor{gray!15}  
        &  4  & $\times$
        &  5  & $\times$ 
        &  5 \cellcolor{gray!15} & $\sqrt{}$\cellcolor{gray!15}
        &  5  & $\times$
        &  5  & $\times$
        &  5 \cellcolor{gray!15} & $\sqrt{}$\cellcolor{gray!15}  \\
~ &q2fsm &  3 & $\sqrt{}$ 
        &  5 & $\sqrt{}$ 
        & 5\cellcolor{gray!15} &$\sqrt{}$  \cellcolor{gray!15}   
        &  5 & $\sqrt{}$ 
        &  5 & $\sqrt{}$  
        & 5\cellcolor{gray!15} &$\sqrt{}$  \cellcolor{gray!15} 
        &  5 & $\sqrt{}$ 
        &  5 & $\sqrt{}$  
        & 5\cellcolor{gray!15} &$\sqrt{}$  \cellcolor{gray!15}   \\

~ &fsm1 & 0 & -
        &  5 & $\sqrt{}$  
        & 5\cellcolor{gray!15} &$\sqrt{}$  \cellcolor{gray!15}
        &  5 & $\sqrt{}$ 
        &  5 & $\sqrt{}$  
        & 5\cellcolor{gray!15} &$\sqrt{}$  \cellcolor{gray!15} 
        &  5 & $\sqrt{}$ 
        &  5 & $\sqrt{}$  
        & 5\cellcolor{gray!15} &$\sqrt{}$  \cellcolor{gray!15}   \\

~ &q2afsm & 0  & -
        &  5 & $\times$
        & 5 \cellcolor{gray!15} & $\sqrt{}$\cellcolor{gray!15}   
        & 5 & $\sqrt{}$
        & 5 & $\sqrt{}$
        & 5 \cellcolor{gray!15} & $\times$  \cellcolor{gray!15}
        & 5 & $\sqrt{}$
        & 5 & $\sqrt{}$
        & 5 \cellcolor{gray!15} & $\times$  \cellcolor{gray!15}  \\
\hhline{~*{19}{-}}
~ &fsm3 $^\dagger$ & 0 & -
        &  0 & -  
        & 5\cellcolor{gray!15} &$\sqrt{}$  \cellcolor{gray!15}  
        & 3 & $\times$ 
        & 1 & $\times$ 
        & 5\cellcolor{gray!15} &$\sqrt{}$  \cellcolor{gray!15}  
        &  5 & $\times$ 
        &  5 & $\times$  
        & 5\cellcolor{gray!15} &$\sqrt{}$  \cellcolor{gray!15}  \\
~ &q2fsm $^\dagger$ &  5 & $\times$ 
        &  5 & $\times$ 
        & 5\cellcolor{gray!15} &$\sqrt{}$  \cellcolor{gray!15} 
        &  5 & $\times$ 
        &  5 & $\times$  
        & 5\cellcolor{gray!15} &$\sqrt{}$  \cellcolor{gray!15} 
        &  5 & $\times$ 
        &  5 & $\times$  
        & 5\cellcolor{gray!15} &$\sqrt{}$  \cellcolor{gray!15}   \\
~ &fsm1s $^\dagger$  &  3 & $\times$ 
        &  5 & $\times$ 
        & 5 \cellcolor{gray!15} & $\sqrt{}$  \cellcolor{gray!15}   
        & 5 & $\times$ 
        & 1 & $\times$ 
        & 5 \cellcolor{gray!15} & $\sqrt{}$  \cellcolor{gray!15}  
        & 5 & $\times$ 
        & 5 & $\times$
        & 5 \cellcolor{gray!15} & $\sqrt{}$  \cellcolor{gray!15}    \\
~ &fsm3s $^\dagger$  & 2  & $\times$
        &  1 & $\times$
        & 5 \cellcolor{gray!15} & $\sqrt{}$  \cellcolor{gray!15}  
        & 5 & $\times$
        & 5 & $\times$
        & 5 \cellcolor{gray!15} & $\sqrt{}$  \cellcolor{gray!15}  
        & 5 & $\times$
        & 5 & $\times$
        & 5 \cellcolor{gray!15} & $\sqrt{}$  \cellcolor{gray!15}    \\
~ &fsmonehot $^\dagger$  & 0  & -
        &  2 & $\times$
        & 0 \cellcolor{gray!15} & - \cellcolor{gray!15}   
        & 0 & -
        & 0 & -
        & 0 \cellcolor{gray!15} & - \cellcolor{gray!15} 
        & 5 & $\times$ 
        & 5 & $\times$
        & 5 \cellcolor{gray!15} & $\sqrt{}$  \cellcolor{gray!15}  \\
~ &fsm2s $^\dagger$  & 0  & -
        &  2 & $\times$
        & 0 \cellcolor{gray!15} & - \cellcolor{gray!15}   
        & 0 & -
        & 1 & $\times$
        & 5 \cellcolor{gray!15} & $\sqrt{}$  \cellcolor{gray!15}
        & 5 & $\times$ 
        & 5 & $\times$
        & 5 \cellcolor{gray!15} & $\sqrt{}$  \cellcolor{gray!15}  \\

~ &q3bfsm$^\dagger$  & 0  & -
        &  5 & $\times$ 
        & 0\cellcolor{gray!15} & -\cellcolor{gray!15}   
        & 4 & $\times$ 
        & 5 & $\times$ 
        & 5 \cellcolor{gray!15} & $\sqrt{}$  \cellcolor{gray!15} 
        & 0 & -
        & 2 & $\times$
        & 5 \cellcolor{gray!15} & $\sqrt{}$  \cellcolor{gray!15}  \\
        
\hhline{~*{19}{-}}

~ &fsm1 $^\triangle$ &  0 & -
        & 5  & $\sqrt{}$
        & 5 \cellcolor{gray!15} & $\sqrt{}$  \cellcolor{gray!15}  
        & 4 & $\sqrt{}$
        & 5  & $\sqrt{} $
        & 5 \cellcolor{gray!15} & $\sqrt{}$  \cellcolor{gray!15} 
        & 5  & $\sqrt{} $ 
        & 5  & $\sqrt{} $
        & 5 \cellcolor{gray!15} & $\sqrt{}$  \cellcolor{gray!15}  \\

~ &fsm1s $^\triangle$ &  4 & $\sqrt{} $
        & 5  & $\sqrt{} $
        & 5 \cellcolor{gray!15} & $\sqrt{}$  \cellcolor{gray!15}      
        & 0 & -
        & 5 & $\sqrt{} $
        & 5 \cellcolor{gray!15} & $\sqrt{}$  \cellcolor{gray!15}   
        & 5 & $\sqrt{} $
        & 5 & $\sqrt{} $ 
        & 5 \cellcolor{gray!15} & $\sqrt{}$  \cellcolor{gray!15}     \\
~ &fsm3 $^\triangle$ & 0  & -
        & 0  & -
        & 5 \cellcolor{gray!15} & $\sqrt{}$  \cellcolor{gray!15}    
        & 0  & -
        & 0  & -
        & 5 \cellcolor{gray!15} & $\sqrt{}$  \cellcolor{gray!15} 
        & 3 & $\sqrt{}$
        & 5 & $\sqrt{}$
        & 5 \cellcolor{gray!15} & $\sqrt{}$  \cellcolor{gray!15}   \\
~ &fsm3s $^\triangle$ & 0  & -
        & 3  & $\times$
        & 5 \cellcolor{gray!15} & $\sqrt{}$  \cellcolor{gray!15} 
        & 1 & $\times$
        & 4 & $\times$
        & 5 \cellcolor{gray!15} & $\sqrt{}$  \cellcolor{gray!15} 
        & 5 & $\times$
        & 5 & $\times$
        & 5 \cellcolor{gray!15} & $\sqrt{}$  \cellcolor{gray!15}  \\

\hline

\multirow{2}{*}{\uppercase\expandafter{\romannumeral2}} &
ARC & 2  & $\times$
        & 5 & $\times$ 
        & 5 \cellcolor{gray!15} & $\sqrt{}$  \cellcolor{gray!15} 
        & 3 & $\times$ 
        & 5 & $\times$ 
        & 5 \cellcolor{gray!15} & $\sqrt{}$  \cellcolor{gray!15}
        & 5 & $\times$ 
        & 5 & $\times$ 
        & 5 \cellcolor{gray!15} & $\sqrt{}$  \cellcolor{gray!15}  \\

~ &ARC $^\triangle$ & 0  & - 
        & 1  & $\sqrt{} $
        & 0\cellcolor{gray!15} & -\cellcolor{gray!15}   
        & 0  & -
        & 0  & -
        & 5 \cellcolor{gray!15} & $\sqrt{}$  \cellcolor{gray!15}  
        & 2 & $\sqrt{} $
        & 0  & - 
        & 0\cellcolor{gray!15} & -\cellcolor{gray!15}  \\
\hline

\multirow{4}{*}{\uppercase\expandafter{\romannumeral3}} &SHA-512   &  0 & -
        &  5 & $\sqrt{}$
        &  5 \cellcolor{gray!15} & $\sqrt{}$\cellcolor{gray!15}   
        &  0 & -
        &  5 & $\sqrt{}$
        &  5 \cellcolor{gray!15} & $\sqrt{}$\cellcolor{gray!15}   
        &  5 & $\sqrt{}$
        &  5 & $\sqrt{}$
        &  5 \cellcolor{gray!15} & $\sqrt{}$\cellcolor{gray!15}    \\
~ &AES     &  0 & -  
        &  0 & - 
        & 0\cellcolor{gray!15} & -\cellcolor{gray!15}   
        &  3 & $\sqrt{}$
        &  3 & $\sqrt{}$
        & 5 \cellcolor{gray!15} &  $\sqrt{}$ \cellcolor{gray!15}  
        &  5 & $\sqrt{}$
        &  5 & $\sqrt{}$ 
        &5 \cellcolor{gray!15} & $\sqrt{}$ \cellcolor{gray!15}  \\
~ &PICO(b)   & 0  & -  
        & 1 & $\times$
        & 0 \cellcolor{gray!15} & - \cellcolor{gray!15}   
        & 0  & -
        & 5 & $\times$ 
        & 5 \cellcolor{gray!15} &  $\times$  \cellcolor{gray!15}  
        & 5 & $\times$  
        & 5 & $\times$   
        & 5\cellcolor{gray!15} &$\times$  \cellcolor{gray!15}  \\

~ &PICO(c) & 0  &  -
        &  5 & $\sqrt{}$ 
        & 5 \cellcolor{gray!15} & $\sqrt{}$ \cellcolor{gray!15}   
        &  5 & $\sqrt{}$  
        &  5 & $\sqrt{}$  
        &  5\cellcolor{gray!15} &   $\sqrt{}$ \cellcolor{gray!15}  
        &  5 & $\sqrt{}$ 
        &  5 & $\sqrt{}$   
        & 5\cellcolor{gray!15} & $\sqrt{}$ \cellcolor{gray!15}  \\

\hline

\multirow{3}{*}{\uppercase\expandafter{\romannumeral4}} &227    &  1 &  $\times$  
        & 5 &  $\times$ 
        & 5 \cellcolor{gray!15} & $\sqrt{}$ \cellcolor{gray!15}   
        &  1 &  $\times$   
        &  0 & -
        & 4 \cellcolor{gray!15} &  $\sqrt{}$ \cellcolor{gray!15}  
        & 5 &  $\times$  
        & 5 &  $\times$   
        & 5 \cellcolor{gray!15} & $\sqrt{}$ \cellcolor{gray!15}   \\

~ &995    & 0  & - 
        & 5 &  $\sqrt{}$
        & 5 \cellcolor{gray!15} &  $\sqrt{}$\cellcolor{gray!15}   
        & 4 &  $\sqrt{}$
        & 5 &  $\sqrt{}$  
        & 5 \cellcolor{gray!15} &  $\sqrt{}$   \cellcolor{gray!15}  
        & 5 &  $\sqrt{}$ 
        & 5 &  $\sqrt{}$ 
        & 5 \cellcolor{gray!15} &  $\sqrt{}$   \cellcolor{gray!15}    \\

~ &601    & 0  & - 
        & 0  & - 
        & 5 \cellcolor{gray!15} & $\times$ \cellcolor{gray!15}   
        & 5 &  $\times$  
        & 4 &  $\times$  
        & 5 \cellcolor{gray!15} & $\times$ \cellcolor{gray!15}  
        & 5 & $\times$ 
        &  5 &  $\times$ 
        & 5\cellcolor{gray!15} & $\times$ \cellcolor{gray!15}  \\
        
\hline

~ &Pass Rate & 16$\%$ &2/25&67$\%$&8/25 &80$\%$&16/25&58$\%$&7/25&70$\%$&9/25&95$\%$ &20/25&92$\%$&10/25&96$\%$&10/25& 96$\%$ & 21/25\\
\hline
\end{tabular}

\begin{tablenotes}
\item[a] 
\uppercase\expandafter{\romannumeral1} : VerilogEval\cite{verilogeval}.
\uppercase\expandafter{\romannumeral2} : ARC-FSM\cite{Arc-fsm-g}.
\uppercase\expandafter{\romannumeral3} : Real industrial FSM examples.
\uppercase\expandafter{\romannumeral3} : Artificial dataset~\cite{saha2024empowering}.
\item[b] Examples marked with special symbols represent injected vulnerabilities. $*$ indicates injection of normal\&coding vulnerabilities(e.g., ``Hamming Distance''). $\dagger$ indicates injection of structural vulnerabilities(e.g., ``Dead State''). $\triangle$ indicates injection of potential vulnerabilities(e.g., ``CWE-190'').
\vspace{-1em}
\end{tablenotes}

\end{threeparttable}
\end{table*}

\section{Evaluation}

This paper evaluates the effectiveness of LLMs in security verilog code generation and examines the impact of our proposed methods. Specifically, we investigated three Research Questions (RQs):
(1) \textit{RQ1: How does SecFSM perform?}
(2) \textit{RQ2: How do different module contribute to the effectiveness of SecFSM?}
(3) \textit{RQ3: How does LLM backbones impact?}
We also perform a case study to demonstrate the effectiveness of our SecFSM.

\subsection{Experiment Setup}
\subsubsection{Datasets.}
To evaluate the effectiveness of the method, we compile a comprehensive test case set comprising academic datasets~\cite{verilogeval}, artifical datasets~\cite{saha2024empowering}, industrial cases~\cite{Arc-fsm-g}, and academic papers~\cite{Arc-fsm-g}. Following the vulnerability injection approach~\cite{saha2024empowering}, we systematically introduced vulnerability into a subset of these cases. The types and quantities of injected vulnerabilities of the cases are described in detail in the appendix. This augmented collection constitutes our final test benchmark for experimental validation.
Figure \ref{fig:dataset} shows the sources of cases used for evaluation.

\subsubsection{Metric.}
We evaluate the performance of the different methods based on two metrics: functionality (``Fun'') and security (``Sec"). The ``Fun" metric denotes the number of functional tests successfully passed across five experimental trials. The ``Sec" metric is assigned a checkmark (``$\sqrt{}$") if at least one security test is passed within the results of the functional testing suite for a given trial; otherwise, a cross (``×") is assigned. In instances where the LLM fails all functional tests in all five trials, ``Sec" metric is designated as not applicable (``-").

\subsubsection{Baselines.}
To evaluate the security of different methods for generating FSM code, we compared three methods: (1) Base, (2) RAG, and (3) SecFSM (the method proposed in this paper). To ensure fairness in the experiment, RAG and SecFSM use the same security knowledge. To eliminate the impact of different LLMs on the experimental results, we selected a total of three LLMs (GPT-4o~\cite{gpt-4o}, Claude 3.5~\cite{claude3.5}, and Deepseeker-R1~\cite{guo2025deepseek}). Additionally, the temperature of all LLMs is set to a lower value (temperature = 0.2).

\subsection{Overall Performance (RQ1)}

Table \ref{Security Evaluation results} shows the pass rates for functionality and security.
From this table, we can see that the results of \textit{Base} performs poorly on security due to insufficient FSM security knowledge. Due to the poor performance of certain models(GPT-4o) in terms of code functionality, this also affects the security of the code.
Furthermore, we can see that \textit{RAG} performs poorly in improving code security due to incomplete security prompts. But it boosts functionality for models with constrained Verilog capabilities (e.g., GPT-4o), while high-capability models like DeepSeek-R1 generate correct code without supplementary support.
Finally, \textit{SecFSM} exhibits marked efficacy in mitigating structural vulnerabilities (notably ``Dead State") and moderate improvements against coding vulnerabilities (``Hamming distance", ``FIF"), substantially increasing secured design instances from 2 to 15 for GPT-4o, 7 to 20 for Claude3.5, and 10 to 21 for DeepSeek-R1 across all experimental groups.

\subsection{Module Performance (RQ2)}

\begin{table*}[!t]
\centering
\small 
\captionsetup{skip=2pt}
\caption{\label{Ablation experiments} Module Performance.}

\setlength{\tabcolsep}{4pt} 
\renewcommand{\arraystretch}{1}
\begin{tabular}{
    l  
    l || cc| cc| cc| cc || cc| cc| cc| cc
}
\hline
&\multirow{3}{*}{\textbf{Case}} 
& \multicolumn{8}{c||}{\textbf{SecFSM(GPT-4o)}} 
& \multicolumn{8}{c}{\textbf{SecFSM(Deepseek-R1)}} \\
\cline{3-18} 
& & \multicolumn{2}{c|}{\textbf{\makecell{Without \\ KR}}} 
& \multicolumn{2}{c|}{\textbf{\makecell{Without \\ PA}}} 
& \multicolumn{2}{c|}{\textbf{\makecell{Without KR \\and PA}}} 
& \multicolumn{2}{c||}{\textbf{\makecell{With KR \\ and PA}}}  
& \multicolumn{2}{c|}{\textbf{\makecell{Without \\ KR}}} 
& \multicolumn{2}{c|}{\textbf{\makecell{Without \\ PA}}} 
& \multicolumn{2}{c|}{\textbf{\makecell{Without KR \\and PA}}} 
& \multicolumn{2}{c}{\textbf{\makecell{With KR and \\ PA}}}  \\ 
\cline{3-18}
& & Fun & Sec & Fun & Sec & Fun & Sec & Fun & Sec  
& Fun & Sec & Fun & Sec & Fun & Sec & Fun & Sec \\ 
\hline
\multirow{15}{*}{I}& q2fsm $^*$ 
 & 5 & $\times$ & 5 & $\times$ & 5 & $\times$ &  5 \cellcolor{gray!15} & $\times$\cellcolor{gray!15}   
 & 5 & $\times$ & 5 & $\sqrt{}$ & 5 & $\times$ & 5 \cellcolor{gray!15} & $\sqrt{}$\cellcolor{gray!15}  \\ 
&fsm3s $^*$
 & 0 & - & 5 & $\times$ & 0 & - &  5 \cellcolor{gray!15} & $\times$\cellcolor{gray!15}   
 & 5 & $\times$ & 5 & $\sqrt{}$ & 5 & $\times$ & 5 \cellcolor{gray!15} & $\sqrt{}$\cellcolor{gray!15} \\
&fsm3s
 & 5 & $\sqrt{}$ & 5 & $\sqrt{}$ & 5 & $\sqrt{}$ &  5 \cellcolor{gray!15} & $\sqrt{}$\cellcolor{gray!15}   
 & 5 & $\sqrt{}$ & 5 & $\sqrt{}$  & 5 & $\sqrt{}$  &  5 \cellcolor{gray!15} & $\sqrt{}$\cellcolor{gray!15}   \\
 
&fsm1
 & 5 & $\sqrt{}$ & 5 & $\sqrt{}$  & 5 & $\sqrt{}$  &  5 \cellcolor{gray!15} & $\sqrt{}$\cellcolor{gray!15}   
 & 5 & $\sqrt{}$ & 5 & $\sqrt{}$  & 5 & $\sqrt{}$  &  5 \cellcolor{gray!15} & $\sqrt{}$\cellcolor{gray!15}   \\
\hhline{~*{17}{-}}
&fsm3 $^\dagger$ 
 & 5 & $\sqrt{}$ & 5 & $\times$  &  5 & $\times$  &  5 \cellcolor{gray!15} & $\sqrt{}$\cellcolor{gray!15}  
 & 5 & $\sqrt{}$ & 5 & $\times$  & 5 & $\times$  &  5 \cellcolor{gray!15} & $\sqrt{}$\cellcolor{gray!15}   \\
&q2fsm $^\dagger$ 
 & 5 & $\sqrt{}$ & 5 & $\times$  &  5 & $\times$  &  5 \cellcolor{gray!15} & $\sqrt{}$\cellcolor{gray!15}  
 & 5 & $\sqrt{}$ & 5 & $\times$  & 5 & $\times$  &  5 \cellcolor{gray!15} & $\sqrt{}$\cellcolor{gray!15}    \\
&fsm1s $^\dagger$   
 & 5 & $\sqrt{}$ & 5 & $\times$  &  5 & $\times$ &  5 \cellcolor{gray!15} & $\sqrt{}$\cellcolor{gray!15}   
 & 5 & $\sqrt{}$ & 5 & $\times$  & 5 & $\times$  &  5 \cellcolor{gray!15} & $\sqrt{}$\cellcolor{gray!15}   \\
&fsm3s $^\dagger$   
 & 5 & $\sqrt{}$ & 5 & $\sqrt{}$  & 5 & $\sqrt{}$  &  5 \cellcolor{gray!15} & $\sqrt{}$\cellcolor{gray!15}   
 & 5 & $\sqrt{}$ & 5 & $\times$  & 5 & $\times$  &  5 \cellcolor{gray!15} & $\sqrt{}$\cellcolor{gray!15}   \\
&fsm2s $^\dagger$  
 &0 &- &0 &- &0 &- &0 \cellcolor{gray!15} & -\cellcolor{gray!15}   
 & 5 & $\sqrt{}$ & 5 & $\times$  & 5 & $\times$  &  5 \cellcolor{gray!15} & $\sqrt{}$\cellcolor{gray!15} \\

&q3bfsm $^\dagger$ 
 & 5 & $\sqrt{}$ & 5 & $\times$  &  5 & $\times$ & 5 \cellcolor{gray!15} & $\sqrt{}$\cellcolor{gray!15}  
 & 5 & $\sqrt{}$ & 5 & $\times$  &  5 & $\times$  & 5 \cellcolor{gray!15} & $\sqrt{}$\cellcolor{gray!15}    \\
\hhline{~*{17}{-}}
&fsm1 $^\triangle$
 & 5 & $\sqrt{}$  & 5 & $\sqrt{}$  &  5 & $\sqrt{}$  & 5 \cellcolor{gray!15} & $\sqrt{}$\cellcolor{gray!15}   
 & 5 & $\sqrt{}$   & 5 & $\sqrt{}$  &  5 & $\sqrt{}$   & 5 \cellcolor{gray!15} &$\sqrt{}$  \cellcolor{gray!15}  \\
&fsm1s $^\triangle$
 &5 & $\sqrt{}$  & 5 & $\sqrt{}$  & 0 & - & 5 \cellcolor{gray!15} & $\sqrt{}$\cellcolor{gray!15}   
 & 5 & $\sqrt{}$   & 5 & $\sqrt{}$  &  5 & $\sqrt{}$   & 5 \cellcolor{gray!15} &$\sqrt{}$  \cellcolor{gray!15}   \\
&fsm3 $^\triangle$
 & 0 & -  & 5 & $\sqrt{}$  & 0 & - & 5 \cellcolor{gray!15} & $\sqrt{}$\cellcolor{gray!15}   
 &  5 & $\sqrt{}$   & 5 & $\sqrt{}$  &  5 & $\sqrt{}$   & 5 \cellcolor{gray!15} &$\sqrt{}$  \cellcolor{gray!15}   \\
&fsm3s $^\triangle$ 
 & 5 & $\times$   & 5 & $\sqrt{}$  & 5 & $\times$ & 5 \cellcolor{gray!15} & $\sqrt{}$\cellcolor{gray!15}  
 & 5 & $\times$  &5 & $\times$  & 5 & $\times$  &  5 \cellcolor{gray!15} &$\sqrt{}$  \cellcolor{gray!15}   \\
\hline

\multirow{1}{*}{\uppercase\expandafter{\romannumeral2}} &SHA-512 
 & 5 & $\sqrt{}$ & 5 & $\sqrt{}$ & 5 & $\sqrt{}$ &  5 \cellcolor{gray!15} & $\sqrt{}$\cellcolor{gray!15}   
 & 5 & $\sqrt{}$ & 5 & $\sqrt{}$  & 5 & $\sqrt{}$  &  5 \cellcolor{gray!15} & $\sqrt{}$\cellcolor{gray!15}   \\
\hline

\multirow{2}{*}{\uppercase\expandafter{\romannumeral3}}& ARC
 & 5 & $\sqrt{}$ & 5 & $\times$  &  5 & $\times$  & 5 \cellcolor{gray!15} & $\sqrt{}$\cellcolor{gray!15}  
 & 5 & $\sqrt{}$ & 5 & $\times$  &  5 & $\times$  & 5 \cellcolor{gray!15} & $\sqrt{}$\cellcolor{gray!15}    \\
&ARC$^\triangle$
 & 0 & - & 5 & $\sqrt{}$  & 0 & - & 0 \cellcolor{gray!15} & -\cellcolor{gray!15}   
 & 0 & - & 0 & - & 0 & - & 0\cellcolor{gray!15} & -\cellcolor{gray!15}  \\
\hline
&Pass Rate 
 &76$\%$&11/17&94$\%$&9/17&76$\%$&5/17&88$\%$&13/17 
 &94$\%$&13/17&94$\%$&8/17&94$\%$&6/17&94$\%$&16/17\\ 
\hline
\end{tabular}
\begin{tablenotes}
\item[a] KR: knowledge retrieval module. PA: pre-analysis module.
\end{tablenotes}
\vspace{-1em}
\end{table*}

To evaluate the contribution of each constituent component within the proposed method, four ablative configurations are implemented:
(1) Without knowledge retrieval module: we remove the FSKG-based security knowledge retrieval module. 
(2) Without pre-analysis module: we remove the pre-analysis module while retaining full FSKG knowledge. 
(3) Without knowledge retrieval module and pre-analysis module: we remove pre-analysis module and knowledge retrieval module. 
(4) Complete SecFSM.
\begin{figure}[!t]
    \centering
    \includegraphics[width=0.4\textwidth]{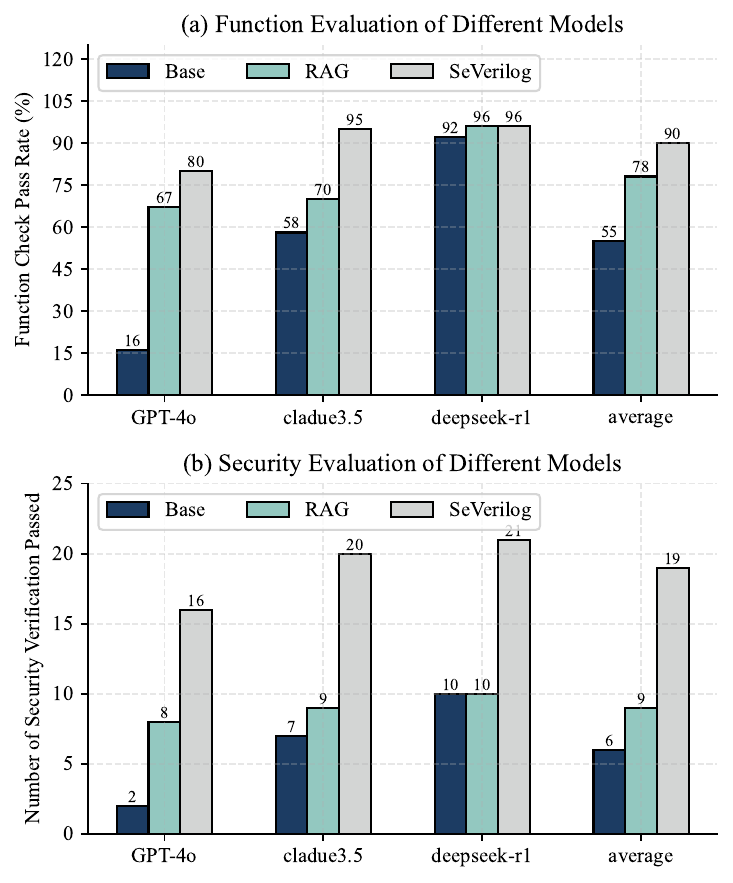}
    \caption{Evaluation of Different Models. Compare three methods (Base, RAG, and SecFSM) using different LLMs (GPT-4o, Claude3.5, and Deepseek-R1).}  
    \vspace{-1em}
    \label{fig:different model}
\end{figure}

Table \ref{Ablation experiments} shows that both the pre-analysis module and the knowledge retrieval module are essential for enhancing FSM code security. 
We can see that the knowledge retrieval module significantly mitigates coding vulnerabilities(such as, q2fsm$^*$). This proves the importance of the security knowledge provided by knowledge retrieval module.
Additionally, we can see that LLMs exhibit limitations in identifying structural vulnerabilities (e.g., "Dead State") through textual analysis alone, and the pre-analysis module's critical role in detecting such vulnerabilities is substantiated from (1).
Furthermore, we can see that LLMs exhibit robust capabilities in processing security-related textual knowledge from (2).
Interestingly, (3) demonstrates that autogen~\cite{wu2023autogen} and the integration of the planning module significantly enhance the functional correctness of code generated by the GPT-4o.

\subsection{LLM Backbone Impact (RQ3)}

Figure \ref{fig:different model} shows the performance of various methods in three LLM backbones on the "Fun" and "Sec" metrics. From this figure, it can be observed that a positive correlation exists between the pass rate and the capacity of the LLM backbones. 
Furthermore, it is evident that the LLM backbone has a significant impact on the functionality of Verilog code. 
However, the impact of the LLM backbone on security is not significant because of the lack of security knowledge in LLM backbones. 
Finally, the results indicate that the ability of LLM to generate Verilog code is very important in order to obtain Verilog code for secure FSM. 
Therefore, efforts to secure FSM generation should focus first on boosting the LLM's accuracy in generating functionally correct Verilog code for FSM.

\subsection{Case Study}
\begin{figure}[!t]
    \centering
    \includegraphics[width=0.42\textwidth]{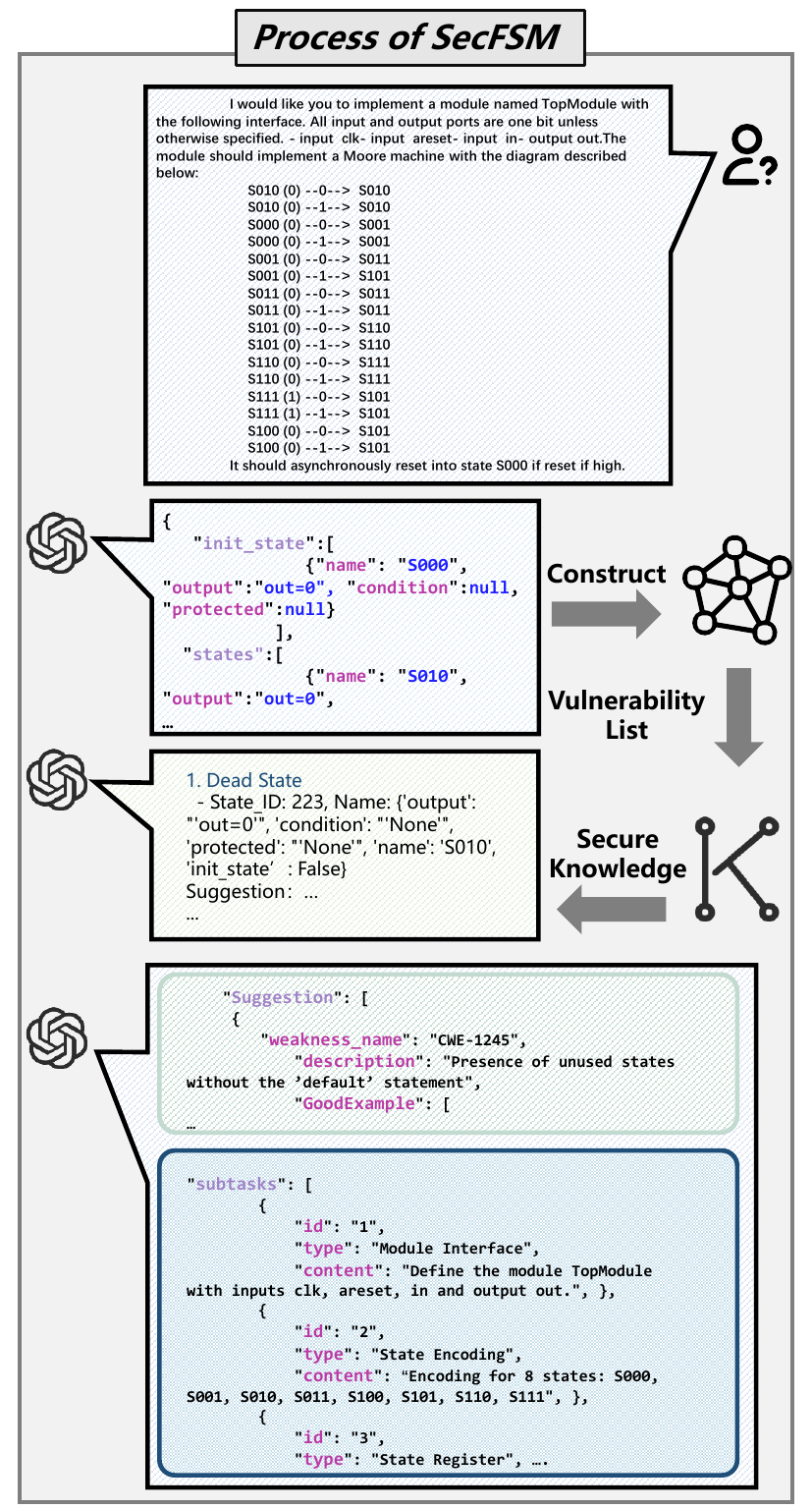}
    \caption{Case Study. We introduced the process from users' requirement to pre-analysis (construction of $G_{ST}$) and knowledge retrieval to security prompt words.}
    \label{fig:Case}
\end{figure}
Figure \ref{fig:Case} depicts the process from user input through pre-analysis and knowledge retrieval to the generation of the prompts which includes security information and task plan.
There are two reasons why our method is effective. (1) We built a domain knowledge graph as the external knowledge base of LLM. It complements LLM's knowledge base. (2) We propose a pre-analysis method based on the structural characteristics of FSM.It can analyze the vulnerabilities of FSM in the requirements phase.

Regarding vulnerabilities associated with coding (e.g., Hamming distance), security knowledge demonstrates efficacy in enhancing code security for scenarios characterized by simple state transitions, as evidenced in cases q2fsm$^*$ and fsm3s$^*$. Conversely, performance degrades significantly in more complex cases (995 and 601), attributable to inadequate comprehension of the state transition logic. The pre-analysis module within SecFSM proves effective for cases where state transition information is explicitly defined. 
However, for cases employing nonstandard state description formats (e.g., q2afsm), the pre-analysis module may extract incorrect information, resulting in misjudgments.

\section{Related Works}
\paragraph{FSM Security.}
The security of FSM has been extensively studied, including vulnerability detection, state encoding optimization, and auxiliary works. In vulnerability detection, AVFSM~\cite{AVFSM} developed a vulnerability analysis framework of FSM to analyze different vulnerabilities from the gate-level netlist. Arc-fsm-g~\cite{Arc-fsm-g} defines a set of security rules. In the state coding, the FSM design process proposed in ~\cite{Security-aware} starts from the security-aware coding scheme, which makes the FSM resilient to fault attacks. In addition, Dipayan Saha~\cite{saha2024empowering} built a vulnerable hardware database. RTL-FSMx~\cite{Rtl-fsmx} can accurately recover all control FSMs from RTL code of different complexities and sizes within seconds. 
\paragraph{RTL Code Security.}
In previous research, there has been work involving RTL code security~\cite{security}. CWEAT~\cite{Don'tCWEAT} investigated vulnerabilities in the hardware CWE database that may be suitable for the identification of source code analysis and proposed a prototype scanner to provide security-related feedback. RTLFixer~\cite{ahmad2024hardware} uses the RAG method and ReAct prompt to use a large language model (LLM) to automatically fix syntax errors in Verilog code.
These approaches tend to detect and mitigate vulnerabilities in code based on RTL code.
To detect the security questions as early as possible, we have introduced an vulnerability pre-analysis method that relies on requirements to analyze structural vulnerabilities and potential vulnerabilities.

\paragraph{LLM for EDA.}
The use of LLM in EDA is an emerging field~\cite{llm4eda}, as evidenced by recent advancements. Chipnemo~\cite{chipnemo} fine-tunes pre-trained LLMs to build a chip-specific assistive chatbot. Smarton-AI~\cite{smart-ai} proposes a new interaction paradigm for utilizing LLM in complex software, including master-slave GPT and question answering GPT.
In the field of RTL evaluation, VerilogEval~\cite{verilogeval} has proposed a comprehensive evaluation dataset consisting of 156 questions. In the field of RTL optimization, BetterV~\cite{betterv} fine-tunes LLMs on domain-specific datasets and combines them with generative discriminators to guide specific PPA requirements. These developments highlight the potential of LLMs to enhance and accelerate RTL design processes. RTLrewrite~\cite{rtlrewriter} uses LLM to optimize RTL code.
\paragraph{RTL Code Generation.}
There are numerous works in the field of RTL code generation. RTLcoder uses the new RTL code dataset and the customized LLM algorithm to implement a lightweight model, with only slight performance degradation. VerilogCoder~\cite{verilogcoder} is a system for Verilog code generation, using the collaborative Verilog tool. To meet the requirements of the professional field, some studies involve the use of specialized tools~\cite{verilogcoder} and methods~\cite{rtlrewriter}.
To address this, we propose a prompt template to guide secure FSM code generation. knowledge retrieval method guided by pre-analysis information has been proposed to accurately complete prompt templates.

\section{Conclusion}
The paper proposes a framework for generating secure Verilog code for FSM design. The core contributions include:
(1) FSM Security Knowledge Graph: vulnerability information, code examples, and mitigation suggestions into a structured knowledge graph to enhance LLM security awareness.
(2) FSM Vulnerability pre-analysis method: analyzing state transition structures against security requirements prior to code generation to identify structural vulnerability and potential vulnerability.
(3) Knowledge retrieval and design a prompt templates: retrieving relevant security knowledge based on pre-analysis results and utilizing security prompt templates to precisely guide LLM code generation. 

\bibliography{aaai2026}

\begin{thebibliography}{29}
\providecommand{\natexlab}[1]{#1}

\bibitem[{Ahmad et~al.(2022)Ahmad, Liu, Collini, Pearce, Fung, Valamehr, Bidmeshki, Sapiecha, Brown, Chakrabarty et~al.}]{Don'tCWEAT}
Ahmad, B.; Liu, W.-K.; Collini, L.; Pearce, H.; Fung, J.~M.; Valamehr, J.; Bidmeshki, M.; Sapiecha, P.; Brown, S.; Chakrabarty, K.; et~al. 2022.
\newblock Don't CWEAT it: Toward CWE analysis techniques in early stages of hardware design.
\newblock In \emph{Proceedings of the 41st IEEE/ACM International Conference on Computer-Aided Design}, 1--9.

\bibitem[{Ahmad et~al.(2024)Ahmad, Thakur, Tan, Karri, and Pearce}]{ahmad2024hardware}
Ahmad, B.; Thakur, S.; Tan, B.; Karri, R.; and Pearce, H. 2024.
\newblock On hardware security bug code fixes by prompting large language models.
\newblock \emph{IEEE Transactions on Information Forensics and Security}.

\bibitem[{Anthropic(2024)}]{claude3.5}
Anthropic, S. 2024.
\newblock Model card addendum: Claude 3.5 haiku and upgraded claude 3.5 sonnet.
\newblock \emph{URL https://api. semanticscholar. org/CorpusID}, 273639283.

\bibitem[{Arifin et~al.(2009)Arifin, Membarth, Amouri, Hannig, and Teich}]{socfsm2019}
Arifin, F.; Membarth, R.; Amouri, A.; Hannig, F.; and Teich, J. 2009.
\newblock FSM-controlled architectures for linear invasion.
\newblock In \emph{2009 17th IFIP International Conference on Very Large Scale Integration (VLSI-SoC)}, 59--64. IEEE.

\bibitem[{Dayarathna, Wen, and Fan(2015)}]{datacenter}
Dayarathna, M.; Wen, Y.; and Fan, R. 2015.
\newblock Data center energy consumption modeling: A survey.
\newblock \emph{IEEE Communications surveys \& tutorials}, 18(1): 732--794.

\bibitem[{Guo et~al.(2025)Guo, Yang, Zhang, Song, Zhang, Xu, Zhu, Ma, Wang, Bi et~al.}]{guo2025deepseek}
Guo, D.; Yang, D.; Zhang, H.; Song, J.; Zhang, R.; Xu, R.; Zhu, Q.; Ma, S.; Wang, P.; Bi, X.; et~al. 2025.
\newblock Deepseek-r1: Incentivizing reasoning capability in llms via reinforcement learning.
\newblock \emph{arXiv preprint arXiv:2501.12948}.

\bibitem[{Han et~al.(2023)Han, Wang, Wang, Yan, and Tian}]{smart-ai}
Han, B.; Wang, X.; Wang, Y.; Yan, J.; and Tian, Y. 2023.
\newblock New interaction paradigm for complex eda software leveraging gpt.
\newblock \emph{arXiv preprint arXiv:2307.14740}.

\bibitem[{He et~al.(2019)He, Guo, Meade, Dutta, Zhao, and Jin}]{socfsmhe}
He, J.; Guo, X.; Meade, T.; Dutta, R.~G.; Zhao, Y.; and Jin, Y. 2019.
\newblock SoC interconnection protection through formal verification.
\newblock \emph{Integration}, 64: 143--151.

\bibitem[{Ho, Ren, and Khailany(2025)}]{verilogcoder}
Ho, C.-T.; Ren, H.; and Khailany, B. 2025.
\newblock Verilogcoder: Autonomous verilog coding agents with graph-based planning and abstract syntax tree (ast)-based waveform tracing tool.
\newblock In \emph{Proceedings of the AAAI Conference on Artificial Intelligence}, volume~39, 300--307.

\bibitem[{Hurst et~al.(2024)Hurst, Lerer, Goucher, Perelman, Ramesh, Clark, Ostrow, Welihinda, Hayes, Radford et~al.}]{gpt-4o}
Hurst, A.; Lerer, A.; Goucher, A.~P.; Perelman, A.; Ramesh, A.; Clark, A.; Ostrow, A.; Welihinda, A.; Hayes, A.; Radford, A.; et~al. 2024.
\newblock Gpt-4o system card.
\newblock \emph{arXiv preprint arXiv:2410.21276}.

\bibitem[{Kibria, Farahmandi, and Tehranipoor(2023)}]{Arc-fsm-g}
Kibria, R.; Farahmandi, F.; and Tehranipoor, M. 2023.
\newblock Arc-fsm-g: Automatic security rule checking for finite state machine at the netlist abstraction.
\newblock In \emph{2023 IEEE International Test Conference (ITC)}, 320--329. IEEE.

\bibitem[{Kibria et~al.(2022)Kibria, Rahman, Farahmandi, and Tehranipoor}]{Rtl-fsmx}
Kibria, R.; Rahman, M.~S.; Farahmandi, F.; and Tehranipoor, M. 2022.
\newblock Rtl-fsmx: Fast and accurate finite state machine extraction at the rtl for security applications.
\newblock In \emph{2022 IEEE International Test Conference (ITC)}, 165--174. IEEE.

\bibitem[{Kocher et~al.(2020)Kocher, Horn, Fogh, Genkin, Gruss, Haas, Hamburg, Lipp, Mangard, Prescher et~al.}]{attack1}
Kocher, P.; Horn, J.; Fogh, A.; Genkin, D.; Gruss, D.; Haas, W.; Hamburg, M.; Lipp, M.; Mangard, S.; Prescher, T.; et~al. 2020.
\newblock Spectre attacks: Exploiting speculative execution.
\newblock \emph{Communications of the ACM}, 63(7): 93--101.

\bibitem[{Lipp et~al.(2020)Lipp, Schwarz, Gruss, Prescher, Haas, Horn, Mangard, Kocher, Genkin, Yarom et~al.}]{attack2}
Lipp, M.; Schwarz, M.; Gruss, D.; Prescher, T.; Haas, W.; Horn, J.; Mangard, S.; Kocher, P.; Genkin, D.; Yarom, Y.; et~al. 2020.
\newblock Meltdown: Reading kernel memory from user space.
\newblock \emph{Communications of the ACM}, 63(6): 46--56.

\bibitem[{Liu et~al.(2023{\natexlab{a}})Liu, Ene, Kirby, Cheng, Pinckney, Liang, Alben, Anand, Banerjee, Bayraktaroglu et~al.}]{chipnemo}
Liu, M.; Ene, T.-D.; Kirby, R.; Cheng, C.; Pinckney, N.; Liang, R.; Alben, J.; Anand, H.; Banerjee, S.; Bayraktaroglu, I.; et~al. 2023{\natexlab{a}}.
\newblock Chipnemo: Domain-adapted llms for chip design.
\newblock \emph{arXiv preprint arXiv:2311.00176}.

\bibitem[{Liu et~al.(2023{\natexlab{b}})Liu, Pinckney, Khailany, and Ren}]{verilogeval}
Liu, M.; Pinckney, N.; Khailany, B.; and Ren, H. 2023{\natexlab{b}}.
\newblock Verilogeval: Evaluating large language models for verilog code generation.
\newblock In \emph{2023 IEEE/ACM International Conference on Computer Aided Design (ICCAD)}, 1--8. IEEE.

\bibitem[{Liu et~al.(2024)Liu, Fang, Lu, Zhang, Zhang, and Xie}]{rtlcoder}
Liu, S.; Fang, W.; Lu, Y.; Zhang, Q.; Zhang, H.; and Xie, Z. 2024.
\newblock Rtlcoder: Outperforming gpt-3.5 in design rtl generation with our open-source dataset and lightweight solution.
\newblock In \emph{2024 IEEE LLM Aided Design Workshop (LAD)}, 1--5. IEEE.

\bibitem[{Nahiyan et~al.(2018)Nahiyan, Farahmandi, Mishra, Forte, and Tehranipoor}]{Security-aware}
Nahiyan, A.; Farahmandi, F.; Mishra, P.; Forte, D.; and Tehranipoor, M. 2018.
\newblock Security-aware FSM design flow for identifying and mitigating vulnerabilities to fault attacks.
\newblock \emph{IEEE Transactions on Computer-aided design of integrated circuits and systems}, 38(6): 1003--1016.

\bibitem[{Nahiyan et~al.(2016)Nahiyan, Xiao, Yang, Jin, Forte, and Tehranipoor}]{AVFSM}
Nahiyan, A.; Xiao, K.; Yang, K.; Jin, Y.; Forte, D.; and Tehranipoor, M. 2016.
\newblock AVFSM: A framework for identifying and mitigating vulnerabilities in FSMs.
\newblock In \emph{Proceedings of the 53rd Annual Design Automation Conference}, 1--6.

\bibitem[{Pei et~al.(2024)Pei, Zhen, Yuan, Huang, and Yu}]{betterv}
Pei, Z.; Zhen, H.-L.; Yuan, M.; Huang, Y.; and Yu, B. 2024.
\newblock Betterv: Controlled verilog generation with discriminative guidance.
\newblock \emph{arXiv preprint arXiv:2402.03375}.

\bibitem[{Peng et~al.(2024)Peng, Zhu, Liu, Bo, Shi, Hong, Zhang, and Tang}]{graph}
Peng, B.; Zhu, Y.; Liu, Y.; Bo, X.; Shi, H.; Hong, C.; Zhang, Y.; and Tang, S. 2024.
\newblock Graph retrieval-augmented generation: A survey.
\newblock \emph{arXiv preprint arXiv:2408.08921}.

\bibitem[{Poehl et~al.(2010)Poehl, Demmerle, Alt, and Obermeir}]{pohone}
Poehl, F.; Demmerle, F.; Alt, J.; and Obermeir, H. 2010.
\newblock Production test challenges for highly integrated mobile phone SOCs—A case study.
\newblock In \emph{2010 15th IEEE European Test Symposium}, 17--22. IEEE.

\bibitem[{Pullini et~al.(2019)Pullini, Rossi, Loi, Tagliavini, and Benini}]{iot}
Pullini, A.; Rossi, D.; Loi, I.; Tagliavini, G.; and Benini, L. 2019.
\newblock Mr. Wolf: An energy-precision scalable parallel ultra low power SoC for IoT edge processing.
\newblock \emph{IEEE Journal of Solid-State Circuits}, 54(7): 1970--1981.

\bibitem[{Saha et~al.(2024)Saha, Yahyaei, Saha, Tehranipoor, and Farahmandi}]{saha2024empowering}
Saha, D.; Yahyaei, K.; Saha, S.~K.; Tehranipoor, M.; and Farahmandi, F. 2024.
\newblock Empowering hardware security with llm: The development of a vulnerable hardware database.
\newblock In \emph{2024 IEEE International Symposium on Hardware Oriented Security and Trust (HOST)}, 233--243. IEEE.

\bibitem[{Thakur et~al.(2024)Thakur, Ahmad, Pearce, Tan, Dolan-Gavitt, Karri, and Garg}]{verigen}
Thakur, S.; Ahmad, B.; Pearce, H.; Tan, B.; Dolan-Gavitt, B.; Karri, R.; and Garg, S. 2024.
\newblock Verigen: A large language model for verilog code generation.
\newblock \emph{ACM Transactions on Design Automation of Electronic Systems}, 29(3): 1--31.

\bibitem[{Wu et~al.(2023)Wu, Bansal, Zhang, Wu, Li, Zhu, Jiang, Zhang, Zhang, Liu et~al.}]{wu2023autogen}
Wu, Q.; Bansal, G.; Zhang, J.; Wu, Y.; Li, B.; Zhu, E.; Jiang, L.; Zhang, X.; Zhang, S.; Liu, J.; et~al. 2023.
\newblock Autogen: Enabling next-gen llm applications via multi-agent conversation.
\newblock \emph{arXiv preprint arXiv:2308.08155}.

\bibitem[{Xie, Zhang, and Peng(2023)}]{security}
Xie, Z.; Zhang, T.; and Peng, Y. 2023.
\newblock Security and Reliability Challenges in Machine Learning for EDA: Latest Advances.
\newblock In \emph{2023 24th International Symposium on Quality Electronic Design (ISQED)}, 1--6. IEEE.

\bibitem[{Yao et~al.(2024)Yao, Wang, Li, Lian, Chen, Chen, Yuan, Xu, and Yu}]{rtlrewriter}
Yao, X.; Wang, Y.; Li, X.; Lian, Y.; Chen, R.; Chen, L.; Yuan, M.; Xu, H.; and Yu, B. 2024.
\newblock Rtlrewriter: Methodologies for large models aided rtl code optimization.
\newblock In \emph{Proceedings of the 43rd IEEE/ACM International Conference on Computer-Aided Design}, 1--7.

\bibitem[{Zhong et~al.(2023)Zhong, Du, Kai, Tang, Xu, Zhen, Hao, Xu, Yuan, and Yan}]{llm4eda}
Zhong, R.; Du, X.; Kai, S.; Tang, Z.; Xu, S.; Zhen, H.-L.; Hao, J.; Xu, Q.; Yuan, M.; and Yan, J. 2023.
\newblock Llm4eda: Emerging progress in large language models for electronic design automation.
\newblock \emph{arXiv preprint arXiv:2401.12224}.

\end{thebibliography}

\end{document}